\documentclass[prl,showpacs,preprintnumbers,twocolumn,amsmath,amssymb,times]{revtex4}

\usepackage{graphicx}
\usepackage{epsfig}

\begin{document}

\newcommand{\bc}{\ensuremath{\mathbf{c}}}
\newcommand{\bj}{\ensuremath{\mathbf{j}}}
\newcommand{\normSt}{\ensuremath{{N}}}
\newcommand{\Kn}{\ensuremath{{\rm Kn}}}

\title{Renormalization of the Lattice Boltzmann Hierarchy}

\author{ Iliya V. Karlin}
\email{karlin@lav.mavt.ethz.ch} \affiliation {Aerothermochemistry
and Combustion Systems Lab, ETH Zurich, 8092 Zurich, Switzerland}

\author{Santosh Ansumali}
\email{ansumali@gmail.com} \affiliation{School of Chemical and
Biomedical Engineering, Nanyang  Technological University, 639798
Singapore, Singapore}

\begin{abstract}
Is it possible to solve Boltzmann-type kinetic equations using
only a small number of particles velocities? We introduce a novel
techniques of solving kinetic equations with (arbitrarily) large
number of particle velocities using only a lattice Boltzmann
method on standard, low-symmetry lattices. The renormalized
kinetic equation is validated with the exact solution of the
planar Couette flow at moderate Knudsen numbers.
\end{abstract}
 \maketitle

The lattice Boltzmann (LB) method has met with a considerable
success  in a wide range of fluid dynamics problems ranging from
turbulent to multi-phase flows \cite{SucciBook}. Recently, much of
the attention was focused on the use of the LB models for
simulation of microflows at moderate Knudsen numbers, $\Kn$, the
ratio of the mean free path to a characteristic flow scale
\cite{AK4,SucciPRL02,AK7,Sofonea05,SucciPRL06,CouettePRL07,Emerson07}.
It is understood by now that LB models form a well-defined
hierarchy \cite{ShanHe,DHT,AK5,ShyamKarlinPRL06}. Each level
$N\ge3$ of the LB hierarchy is characterized by a thoughtfully
chosen set of discrete velocities whose number scales as $N^D$,
where $D$ is the spatial dimension. With increasing level $N$, the
LB hierarchy constitutes a novel approximation of the classical
kinetic theory and has to be considered as an alternative to more
traditional approaches such as higher-order hydrodynamics (Burnett
or super-Burnett \cite{ChapmanCowling}) or Grad's moment systems
\cite{GradH}. One salient feature of the LB hierarchy, which is
crucial for any realistic application and which distinguishes it
from traditional approaches, is that LB is equipped with relevant
boundary conditions derived directly from Maxwell-Boltzmann's
theory \cite{AK4}. However, proceeding to higher levels $N$ (a
must in microflow applications) constitutes an increasingly
difficult computational problem.

In this Letter, we solve the problem of simulating LB models with
large velocity sets on small lattices without sacrificing any
physics or accuracy. The first step in this direction is to
realize that lower-order models  are nothing but closures within
the higher-order models. This simple yet important observation
enables us to formulate the renormalized LB models of the lower
levels in such a way that the additional physics of the
higher-order models is incorporated. In particular, we show with
an example of exact solution in the stationary Couette flow that
the accuracy of the most commonly used planar $D2Q9$ LB can be
enhanced drastically, without introducing additional velocities.
Thus, we introduce a new way of increasing the accuracy of the LB
models without significantly increasing the computational cost.
The methodology developed herein can be used to renormalize other
computational kinetic theories.

We consider the isothermal LB hierarchy of kinetic equations
\begin{equation}
\label{LBM}
\partial_t f_i+ c_{i\alpha}\partial_{\alpha}f_i = Q_i(f),
\end{equation}
where $f_i$ are populations of discrete velocities $c_i$,
$i=1,\dots,N^D$, summation convention is assumed, and $Q$ is the
collision integral satisfying local conservation of density and
momentum, and vanishing at the equilibrium $f^{\rm eq}$, where
\begin{align}
\label{TQA}
\begin{split}
  f_i^{\rm eq}=w_i\left(\rho  + \frac{j_{\alpha}c_{i\, \alpha}}{c_{\rm s}^2}
  +
\frac{j_{\alpha} \,j_{\beta }}{ 2\rho c_{\rm s}^4}
    \left(c_{i\alpha} c_{i\beta} -  c_{\rm s}^2  \delta_{\alpha \beta }\right)
   \right).
\end{split}
\end{align}
Here $\rho=\sum_{i=1}^{N^D}f_i$ is the density,
$j_{\alpha}=\sum_{i=1}^{N^D} c_{i\alpha}f_i$ is the momentum
density, $c_{\rm s}$ is the speed of sound, and we shall use units
in which $c_{\rm s}=1$. The weights $w_i$ and the velocities $c_i$
are so chosen that at each level $N$ the hydrodynamic limit of the
kinetic equation (\ref{LBM}) at low Mach numbers is the
Navier-Stokes equation. While the hydrodynamic limit of all the
models (\ref{LBM}) is the same at each level $N$, their behavior
is markedly different when exploring the micro-flow domain. Our
goal is to modify the lowest-level kinetic equations (\ref{LBM})
in such a way that the non-hydrodynamic features of the
higher-level models are correctly captured by lower-order models.

In order to introduce the main ideas, we consider the
one-dimensional case. For $D=1$, the lowest-order ($N=3$) model
with three velocities $\{-\sqrt{3},0,\sqrt{3}\}$ ($D1Q3$) and
collision integral in the Bhatnagar-Gross-Krook (BGK) form,
$Q_i=(f_i^{\rm eq}-f_i)/\tau$, with a relaxation time $\tau$, can
be written as a moment system for $\rho$, $j$ and pressure
$P=\sum_{i=1}^3c_i^2f_i$:
\begin{align}
\begin{split}
\partial_t\rho &=-\partial_x j,\\
\partial_t j &=-\partial_x P,\\
\partial_t P &=-3\partial_x j -\frac{1}{\tau}(P-P^{\rm eq}),\label{D1Q3}
\end{split}
\end{align}
where $P^{\rm eq}=\rho+ (j^2/\rho)$ is the equilibrium value of
the pressure. Note that when writing equation for the pressure we
have used identity for the energy flux,
$q=\sum_{i=1}^3c_i^3f_i=3j$, which appears as a consequence of a
lattice constraint, $c_i^3=3c_i$. The next ($N=4$) member of the
LB hierarchy is an off-lattice four-velocity model based on the
roots of the fourth-order Hermite polynomial $\{\pm a,\pm b\}$,
where $a=\sqrt{3-\sqrt{6}}$ and $b=\sqrt{3+\sqrt{6}}$ ($D1Q4$).
Assuming a multi-relaxation time collision integral, the
corresponding moment system reads:
\begin{align}
\begin{split}
\partial_t\rho &=-\partial_x j,\\
\partial_t j &=-\partial_x P,\\
\partial_t P &=-\partial_x q-\frac{1}{\tau}(P-P^{\rm eq}),\\
\partial_t q &=-\partial_x (\alpha P+\beta \rho)-\frac{1}{\theta}(q-q^{\rm
eq}), \label{D1Q4}
\end{split}
\end{align}
where $\alpha=\frac{b^4-a^4}{b^2-a^2}=6$,
$\beta=\frac{a^4b^2-b^4a^2}{b^2-a^2}=-3$ are constants of the
four-velocity set, and $q^{\rm eq}=3j$ is equilibrium value of the
energy flux. We have introduced two relaxation times, $\tau$ and
$\theta$, in order to distinguish the relaxation of $P$ and $q$.
The moment system can be realized, for example, as a
quasi-equilibrium kinetic equation \cite{GK94b,Arcidiacono06} with
two relaxation times. Note that the equations for $\{\rho,j,P\}$
are not closed within the system (\ref{D1Q4}).

Now it is easy to see that the $D1Q3$ model (\ref{D1Q3})  is a
closure of the $D1Q4$ moment system (\ref{D1Q4}). Indeed, assuming
$\theta\ll\tau$, and substituting $q\approx q^{(0)}=q^{\rm eq}$
into the equation for pressure, one arrives at a closed sub-system
for $\{\rho,j,P\}$ which is identical to (\ref{D1Q3}). Note that,
from this new angle of view, the aforementioned identity for the
energy flux, $q=3j$ in (\ref{D1Q3}), appears not as a consequence
of the lattice constraint but rather as an implication of the
closure.

Upon realizing this relation between the higher- and lower-level
``bare" kinetic equations (\ref{LBM}), it is tempting to seek
improvements for the closure. The simplest way to do this is to
rescale the time with $\tau$, introduce a smallness parameter
$\eta=\theta/\tau$, and compute the first correction, so that
$q=q^{(0)}+q^{(1)}$, where
\begin{equation}
q^{(1)}=-\tau
\eta\beta\partial_x\rho+\tau\eta(3-\alpha)\partial_xP.
\label{naive}
\end{equation}
This is certainly in the spirit of the Chapman-Enskog method
although note that the system to be closed does not consist solely
of  local conservation laws but also includes relaxation. The
resulting moment system is equivalent to a renormalized kinetic
equation,
\begin{equation}
\partial_{t}f_i+c_i\partial_x f_i -\lambda_i\tau
\eta\partial^2_x[(\alpha-3)P+\beta\rho]
=-\frac{1}{\tau}(f_i-f_i^{\rm eq}), \label{source}
\end{equation}
with $\lambda_{\mp}=1/6$ and $\lambda_0=-1/3$. Thus, we can
realize the one-step renormalization (OSR) (\ref{naive}) (or any
other) as a kinetic equation for populations on the three-velocity
lattice supplemented with a source term. The discrete velocities
of the renormalized kinetic equation (\ref{source}) are on the
lattice, so that the discretization in time and space is
straightforward (see, e.\ g., \cite{HeChenDoolen98}). This simple
example already demonstrates the advantages of the renormalized
lattice kinetic equations.

Now we shall apply the one-step renormalization to the
particularly important two-dimensional sixteen-velocity model
($D2Q16$, $N=4$). The $D2Q16$ model is a tensor product of the two
copies of the $D1Q4$ model considered above, and it (or its
analogs) has attracted attention recently
\cite{CouettePRL07,Szalmas07,Zhang06} as the first LB model which
is capable of describing correctly the transient Knudsen regime,
unlike the standard nine-velocity $D2Q9$ ($N=3$) LB model.

The set of sixteen moments describing the $D2Q16$ model is split
into the locally conserved ($C$), slow relaxing ($S_{\tau}$) and
fast relaxing ($F_{\theta}$) subsystems
\begin{eqnarray}
C&=&\{\rho,j_{x},j_{y}\},\label{Conserved}\\
S_{\tau}&=&\{P_{xx},P_{yy},P_{xy},Q_{xyy},Q_{yxx},\psi\},\label{Slow}\\
F_{\theta}&=&\{Q_{xxx}, Q_{yyy}, \psi_x,
\psi_y,L_{x},L_{y},\phi\}, \label{Fast}
\end{eqnarray}
where ($\langle \varsigma\rangle=\sum_{i=1}^{16} \varsigma_if_i$)
      \[
      \begin{array}{ll}
        P_{\alpha \beta}=\langle c_{\alpha}c_{\beta}\rangle, & Q_{\alpha \beta \gamma}=\langle c_{\alpha} c_{\beta} c_{\gamma}
      \rangle, \\
        \psi=\langle( c_{x}^2 - 1)( c_{y}^2 - 1)\rangle, & \psi_\alpha = \langle(c_{\alpha}^2-3)c_{x}c_{y}\rangle, \\
        L_{\alpha}=\langle c_{\alpha}(c_{x}^2 - 3)(c_{y}^2
      -3)\rangle, & \phi=\langle c_{x} c_{y}(c_{x}^2 - 3)(c_{y}^2 -3)\rangle. \\
      \end{array}
      \]
The closure of the fast subsystem (\ref{Fast}),
$F_{\theta}^{(0)}=F^{\rm eq}$, where
\begin{equation}
F_{\theta}^{(0)}=\{3j_x, 3j_y, 0, 0,0,0,0\},
\end{equation}
renders the moment sub-system for the nine moments $C$ and
$S_{\tau}$ equivalent to the moment system of the $D2Q9$ model
\cite{SucciBook}.
 Thus, again, the standard LBGK
model on the nine-velocity lattice appears as a closure of the
higher-level theory. The one-step renormalization
$F_{\theta}^{(1)}$ is found to be (cf.\ (\ref{naive})):
\begin{align}
\begin{split}
Q_{\alpha\alpha\alpha}^{(1)}&=3\tau\eta(\partial_{\alpha}\rho-\partial_{\alpha}P_{\alpha\alpha}),\\
\psi_{x}^{(1)}&=-3\tau\eta\partial_x(Q_{yxx}-j_y),\\
\psi_{y}^{(1)}&=-3\tau\eta\partial_y(Q_{xyy}-j_x),\\
L_{\alpha}^{(1)}&= -3\tau\eta\partial_{\alpha}\psi,\\
\phi^{(1)}&=0. \label{Correction9}
\end{split}
\end{align}
With (\ref{Correction9}), it is straightforward to write down the
renormalized $D2Q9$ kinetic equation (cf. (\ref{source})) and to
implement the space-time discretization. We do not address this
here. Instead, in order to clearly see the implication of the
one-step renormalization (\ref{Correction9}), we consider the
exact solution of the renormalized $D2Q9$ system in the stationary
Couette flow, where a fluid is enclosed between two parallel
plates separated by a distance $L$. The bottom plate at $y=-L/2$
moves with the velocity $U_1$ and top plate at $y=L/2$ moves with
the velocity $U_2$. The solution of the renormalized model
proceeds essentially along the lines of \cite{CouettePRL07}:
First, the steady-state OSR $D2Q9$ moment system is integrated
under the assumption of unidirectional flow and no mass flux
through the walls. Second, the boundary conditions are applied to
compute the integration constants of the solution. This step is
particularly important: The boundary conditions for the OSR $D2Q9$
model are {\it induced} by the boundary conditions of the $D2Q16$
model. Namely, when the diffusive wall boundary conditions
\cite{AK4} are applied to the $D2Q16$ model, the result is
presented in terms of all the moments $C$, $S_{\tau}$ and
$F_{\theta}$. We then replace, $F_{\theta}\to
F_{\theta}^{(0)}+F_{\theta}^{(1)}$, and obtain the boundary
conditions for the OSR $D2Q9$ system in terms of the $C$ and
$S_\tau$ moments. Application of the boundary condition completes
the solution. Let us introduce the mean free path $l=\sqrt{3}\tau
c_{\rm s}$ and the Knudsen number $\Kn = l/L$. The $x$-component
of the velocity as predicted by the OSR $D2Q9$ model for any
$\eta=\theta/\tau$ is:
\begin{equation}
u_x(y) =A\sinh{\left(\frac{y}{ \Kn \sqrt{\eta}L}\right)}\Delta U +
B\left(\frac{ y}{ L}\right)\Delta U +U, \label{VelocityResult2}
\end{equation}
where $\Delta U=U_2-U_1$ is the relative velocity of the plates,
 $U=(U_1+U_2)/2$ is the centerline velocity, and $A$ and $B$ are constants which depend only
on $\Kn$ and $\eta$:

\begin{align}
\label{Slope16}
\begin{split}
B&=\frac{\mu\sqrt{\eta}
   +  2\sqrt{3} \, \tanh
   \left(\frac{1}{2\sqrt{\eta}
   \Kn}\right)}{
   (4\Kn+\mu)\sqrt{\eta}+
   2(\mu\Kn+\sqrt{3})\tanh \left(\frac{1}{2\sqrt{\eta}
   \Kn}\right)},\\
A&=\frac{4\Kn B}{ \mu^2\sqrt{\eta} \cosh
   \left(\frac{1}{2\sqrt{\eta}
   \Kn}\right)
   +  2\sqrt{3}\mu \sinh
   \left(\frac{1}{2\sqrt{\eta}
   \Kn}\right)},
\end{split}
\end{align}
and $\mu=a+b\approx 3.076$.

It is striking that for $\eta=1$ ($\theta=\tau$), the result
(\ref{VelocityResult2}) and (\ref{Slope16}) becomes {\it
identical} to the one obtained in \cite{CouettePRL07} for the BGK
$D2Q16$ model. We remind (see \cite{Sofonea05,CouettePRL07}) that
the bare $D2Q9$ model predicts only a linear velocity profile in
this problem, $u_x=(1+ 2\Kn)^{-1}(y/L)\Delta U+U$, stripped of the
nonlinear Knudsen layer at the walls. Quite on the contrary, the
renormalized $D2Q9$ model shows clearly the Knudsen layer (first
term in (\ref{VelocityResult2})), which is exactly the same as in
the $D2Q16$ model itself. The reason for this can be traced to the
fact that the renormalization removes the lattice constraint
pertinent to the bare $D2Q9$ model, namely
$Q_{\alpha\alpha\alpha}=3j_{\alpha}$. In the present approach,
this constraint is recognized as a closure relation
$F_{\theta}^{(0)}$ which is then corrected by the first term in
(\ref{Correction9}). Thus, the sense of the renormalization is to
dress the bare kinetic equations with non-hydrodynamic modes so
that they reveal correct behavior at non-vanishing $\Kn$. This is
indeed much in spirit of the renormalization group method
\cite{WilsonKogut} for spin-lattice models where renormalization
improves on the mean-field approximation to dress it with
correlations.

In Fig.\ \ref{Fig1}, the value of the velocity slip at the wall
resulting from (\ref{VelocityResult2}) is compared at various
$\Kn$ with the classical data of Willis \cite{Willis62} for the
linearized Boltzmann-BGK equation, and with results obtained with
the Direct Simulation Monte Carlo (DSMC) method \cite{Bird}. The
result for the bare $D2Q9$ model is also plotted for comparison.
It is clear that the agreement for the renormalized $D2Q9$ model
remains excellent for large values of $\Kn$, and the
renormalization leads to a drastic  improvement of the bare $D2Q9$
model. We conclude this Letter with a number of comments using
again the simple $D1Q4$ model for the sake of argument:

\begin{center}
\begin{figure}[t]
 \includegraphics[scale=0.4]{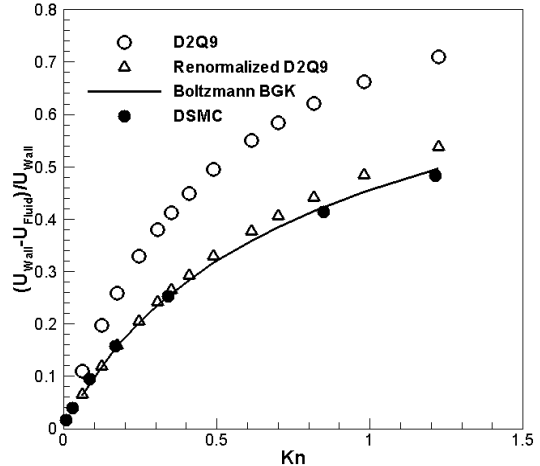}
 \caption{Slip
velocity at the wall as a function of Knudsen number. Line:
linearized Boltzmann-BGK model \cite{Willis62}; Filled circles:
DSMC simulation; Open circles: Standard (bare) $D2Q9$ model
\cite{Sofonea05,CouettePRL07}; Open triangles: One-step
renormalized $D2Q9$ model (\ref{VelocityResult2}), $\eta=1$.
 \label{Fig1}}
 \end{figure}
 \end{center}

 \noindent (i)  The physical meaning of the
renormalization
 is to establish an intermediate level between kinetics and hydrodynamics.
 This intermediate level
happens when the dynamics of $q$ becomes slaved by the dynamics of
$\{\rho, j, P\}$ but the dynamics of $P$ is not yet slaved by the
dynamics of $\{\rho, j\}$. The hydrodynamic limit of model
(\ref{D1Q4}) assumes two smallness parameters, $\epsilon=\tau/T$
and $\mu=\theta/T$ where $T$ is a flow time scale. Instead, we
rearrange it in terms of two other parameters, $\epsilon$ and
$\eta=\theta/\tau=\mu/\epsilon$. Note that $\eta$ need not be
small.

\noindent (ii) Although the simple one-step renormalization is
quite reliable, a rigorous approach to the non-perturbative
renormalization can be based on the invariance equation
\cite{GKbook},
\begin{equation}
\Delta(q)=\partial_t q -\left(\frac{\partial q}{\partial
\rho}\partial_t \rho+ \frac{\partial q}{\partial j}\partial_t
j+\frac{\partial q}{\partial P}\partial_t P\right)=0. \label{inv}
\end{equation}
A stable fixed point of (\ref{inv}) is a fully renormalized $q$.
Owing to a specific feature of the LB hierarchy (linearity of
propagation), a way to solve Eq. (\ref{inv}) (and similar
equations in higher dimensions) is the following: (a) Neglecting
the nonlinearity in $P^{\rm eq}$, we note that the solution
$q^{\rm lin}$ of (\ref{inv}) can be found exactly, following the
route of exact summation of the Chapman-Enskog expansion
\cite{GorbanPRL96,Karlin02}. (b) Once the renormalized linear
closure $q^{\rm lin}$ is obtained, it can be refined to take into
account the nonlinearities. Substituting $q^{\rm lin}$ into
(\ref{inv}), we compute the defect of invariance $\Delta^{\rm
lin}=\Delta(q^{\rm lin})$. With this, a refinement can be written,
$q\approx q^{\rm lin}+a\Delta^{\rm lin}$, where $a$ can be
estimated via a relaxation method \cite{GKbook}.

\noindent (iii) Importantly, the simple OSR or non-perturbative
linear renormalization  should be sufficient for most of the cases
of interest in microflow simulations. In fact, the nonlinearity of
$P^{\rm eq}$ is mainly responsible for the hydrodynamic behavior
of the model (advection term in the Navier-Stokes equations),
whereas the task of renormalization is to remove lattice
constraints and restore such features as Knudsen layers, slip
velocity etc. With this, the renormalized kinetic equations
retaining the full $P^{\rm eq}$ are still nonlinear, as in the
case of Couette flow considered above.

\noindent (iv) As a final remark, in the standard kinetic theory,
the one-step renormalization was first introduced in
\cite{Karlin98} as a correction to Grad's moment systems, and
received some attention after the work \cite{Struchtrup03}.
However, this approach cannot compete with the LB method both in
terms  of computational efficiency and (more restrictively)
because of the lack of well-defined boundary conditions.

In conclusion, the traditional viewpoint on the LB hierarchy
treats each level separately, without any relation across the
levels. Here, an alternative viewpoint is suggested according to
which  bare kinetic equations of the form (\ref{LBM}) on the lower
and computationally attractive levels appear as closures of the
higher-level kinetic equations. Based on this, we suggested to
renormalize the low-order LB equations in such a way that the
physics beyond the standard hydrodynamics is correctly reported
from the higher levels to the lower levels. We demonstrated
analytically that the renormalized lattice Boltzmann model on a
standard velocity set reproduces the Knudsen layer in the Couette
flow which otherwise is possible only with the higher-level
models. In this sense, the renormalized kinetic equations on
standard lattices are {\it the} LB equations, and not the bare
ones, written by a plain analogy with kinetic theory. We note that
the renormalization discussed in this Letter concerns propagation
of non-hydrodynamic effects down the LB hierarchy and not a
renormalization or sub-grid modeling of Navier-Stokes' turbulence,
as in \cite{SCI,AK8}. We are, however, optimistic that the present
methods can also be useful in the latter problem.

We thank S.\ Arcidiacono for the DSMC simulation data of Fig.\
\ref{Fig1}. Support by BFE Project 100862 and CCEM-CH (I.V.K.),
and by NTU Office of Research (S.A.) is gratefully acknowledged.
Part of this work was done during a visit of S.A. to ETH Zurich,
and we gratefully acknowledge the support by the ERCOFTAC.

\bibliography{RKA}

\begin{thebibliography}{29}
\expandafter\ifx\csname natexlab\endcsname\relax\def\natexlab#1{#1}\fi
\expandafter\ifx\csname bibnamefont\endcsname\relax
  \def\bibnamefont#1{#1}\fi
\expandafter\ifx\csname bibfnamefont\endcsname\relax
  \def\bibfnamefont#1{#1}\fi
\expandafter\ifx\csname citenamefont\endcsname\relax
  \def\citenamefont#1{#1}\fi
\expandafter\ifx\csname url\endcsname\relax
  \def\url#1{\texttt{#1}}\fi
\expandafter\ifx\csname urlprefix\endcsname\relax\def\urlprefix{URL }\fi
\providecommand{\bibinfo}[2]{#2}
\providecommand{\eprint}[2][]{\url{#2}}

\bibitem[{\citenamefont{Succi}(2001)}]{SucciBook}
\bibinfo{author}{\bibfnamefont{S.}~\bibnamefont{Succi}},
  \emph{\bibinfo{title}{The {L}attice {B}oltzmann {E}quation for {F}luid
  {D}ynamics and {B}eyond}} (\bibinfo{publisher}{Oxford University Press,
  Oxford}, \bibinfo{year}{2001}).

\bibitem[{\citenamefont{Ansumali and Karlin}(2002)}]{AK4}
\bibinfo{author}{\bibfnamefont{S.}~\bibnamefont{Ansumali}} \bibnamefont{and}
  \bibinfo{author}{\bibfnamefont{I.~V.} \bibnamefont{Karlin}},
  \bibinfo{journal}{Phys.\ Rev.\ E} \textbf{\bibinfo{volume}{66}},
  \bibinfo{pages}{026311} (\bibinfo{year}{2002}).

\bibitem[{\citenamefont{Succi}(2002)}]{SucciPRL02}
\bibinfo{author}{\bibfnamefont{S.}~\bibnamefont{Succi}},
  \bibinfo{journal}{Phys. Rev. Lett.} \textbf{\bibinfo{volume}{89}},
  \bibinfo{pages}{064502} (\bibinfo{year}{2002}).

\bibitem[{\citenamefont{Ansumali and Karlin}(2005)}]{AK7}
\bibinfo{author}{\bibfnamefont{S.}~\bibnamefont{Ansumali}} \bibnamefont{and}
  \bibinfo{author}{\bibfnamefont{I.~V.} \bibnamefont{Karlin}},
  \bibinfo{journal}{Phys. Rev. Lett.} \textbf{\bibinfo{volume}{95}},
  \bibinfo{pages}{260605} (\bibinfo{year}{2005}).

\bibitem[{\citenamefont{Sofonea and Sekerka}(2005)}]{Sofonea05}
\bibinfo{author}{\bibfnamefont{V.}~\bibnamefont{Sofonea}} \bibnamefont{and}
  \bibinfo{author}{\bibfnamefont{R.}~\bibnamefont{Sekerka}},
  \bibinfo{journal}{J. Comput. Phys.} \textbf{\bibinfo{volume}{207}},
  \bibinfo{pages}{639} (\bibinfo{year}{2005}).

\bibitem[{\citenamefont{Horbach and Succi}(2006)}]{SucciPRL06}
\bibinfo{author}{\bibfnamefont{J.}~\bibnamefont{Horbach}} \bibnamefont{and}
  \bibinfo{author}{\bibfnamefont{S.}~\bibnamefont{Succi}},
  \bibinfo{journal}{Phys. Rev. Lett.} \textbf{\bibinfo{volume}{96}},
  \bibinfo{pages}{224503} (\bibinfo{year}{2006}).

\bibitem[{\citenamefont{Ansumali et~al.}(2007)\citenamefont{Ansumali, Karlin,
  Arcidiacono, Abbas, and Prasianakis}}]{CouettePRL07}
\bibinfo{author}{\bibfnamefont{S.}~\bibnamefont{Ansumali}},
  \bibinfo{author}{\bibfnamefont{I.~V.} \bibnamefont{Karlin}},
  \bibinfo{author}{\bibfnamefont{S.}~\bibnamefont{Arcidiacono}},
  \bibinfo{author}{\bibfnamefont{A.}~\bibnamefont{Abbas}}, \bibnamefont{and}
  \bibinfo{author}{\bibfnamefont{N.}~\bibnamefont{Prasianakis}},
  \bibinfo{journal}{Phys. Rev. Lett.} \textbf{\bibinfo{volume}{98}}
  (\bibinfo{year}{2007}).

\bibitem[{\citenamefont{Zhang et~al.}(2007)\citenamefont{Zhang, Gu, Barber, and
  Emerson}}]{Emerson07}
\bibinfo{author}{\bibfnamefont{Y.-H.} \bibnamefont{Zhang}},
  \bibinfo{author}{\bibfnamefont{X.~J.} \bibnamefont{Gu}},
  \bibinfo{author}{\bibfnamefont{R.~W.} \bibnamefont{Barber}},
  \bibnamefont{and} \bibinfo{author}{\bibfnamefont{D.~R.}
  \bibnamefont{Emerson}}, \bibinfo{journal}{Europhys. Lett.}
  \textbf{\bibinfo{volume}{77}}, \bibinfo{pages}{30003} (\bibinfo{year}{2007}).

\bibitem[{\citenamefont{Shan and He}(1998)}]{ShanHe}
\bibinfo{author}{\bibfnamefont{X.}~\bibnamefont{Shan}} \bibnamefont{and}
  \bibinfo{author}{\bibfnamefont{X.}~\bibnamefont{He}},
  \bibinfo{journal}{Phys.\ Rev. Lett.} \textbf{\bibinfo{volume}{80}},
  \bibinfo{pages}{65} (\bibinfo{year}{1998}).

\bibitem[{\citenamefont{Karlin et~al.}(1999)\citenamefont{Karlin, Ferrante, and
  \"{O}ttinger}}]{DHT}
\bibinfo{author}{\bibfnamefont{I.~V.} \bibnamefont{Karlin}},
  \bibinfo{author}{\bibfnamefont{A.}~\bibnamefont{Ferrante}}, \bibnamefont{and}
  \bibinfo{author}{\bibfnamefont{H.~C.} \bibnamefont{\"{O}ttinger}},
  \bibinfo{journal}{Europhys.\ Lett.} \textbf{\bibinfo{volume}{47}},
  \bibinfo{pages}{182} (\bibinfo{year}{1999}).

\bibitem[{\citenamefont{Ansumali et~al.}(2003)\citenamefont{Ansumali, Karlin,
  and \"{O}ttinger}}]{AK5}
\bibinfo{author}{\bibfnamefont{S.}~\bibnamefont{Ansumali}},
  \bibinfo{author}{\bibfnamefont{I.~V.} \bibnamefont{Karlin}},
  \bibnamefont{and} \bibinfo{author}{\bibfnamefont{H.~C.}
  \bibnamefont{\"{O}ttinger}}, \bibinfo{journal}{Europhys. Lett.}
  \textbf{\bibinfo{volume}{63}}, \bibinfo{pages}{798} (\bibinfo{year}{2003}).

\bibitem[{\citenamefont{Chikatamarla and Karlin}(2006)}]{ShyamKarlinPRL06}
\bibinfo{author}{\bibfnamefont{S.~S.} \bibnamefont{Chikatamarla}}
  \bibnamefont{and} \bibinfo{author}{\bibfnamefont{I.~V.}
  \bibnamefont{Karlin}}, \bibinfo{journal}{Phys. Rev. Lett.}
  \textbf{\bibinfo{volume}{97}}, \bibinfo{pages}{190601}
  (\bibinfo{year}{2006}).

\bibitem[{\citenamefont{Chapman and Cowling}(1970)}]{ChapmanCowling}
\bibinfo{author}{\bibfnamefont{S.}~\bibnamefont{Chapman}} \bibnamefont{and}
  \bibinfo{author}{\bibfnamefont{T.~G.} \bibnamefont{Cowling}},
  \emph{\bibinfo{title}{The {M}athematical {T}heory of {N}on-{U}niform
  {G}ases}} (\bibinfo{publisher}{Cambridge University Press, Cambridge},
  \bibinfo{year}{1970}).

\bibitem[{\citenamefont{Grad}(1949)}]{GradH}
\bibinfo{author}{\bibfnamefont{H.}~\bibnamefont{Grad}},
  \bibinfo{journal}{Comm.\ Pure Appl.\ Math.} \textbf{\bibinfo{volume}{2}},
  \bibinfo{pages}{331} (\bibinfo{year}{1949}).

\bibitem[{\citenamefont{Gorban and Karlin}(1994)}]{GK94b}
\bibinfo{author}{\bibfnamefont{A.~N.} \bibnamefont{Gorban}} \bibnamefont{and}
  \bibinfo{author}{\bibfnamefont{I.~V.} \bibnamefont{Karlin}},
  \bibinfo{journal}{Physica A} \textbf{\bibinfo{volume}{206}},
  \bibinfo{pages}{401} (\bibinfo{year}{1994}).

\bibitem[{\citenamefont{Arcidiacono et~al.}(2006)\citenamefont{Arcidiacono,
  Mantzaras, Ansumali, Karlin, Frouzakis, and Boulouchos}}]{Arcidiacono06}
\bibinfo{author}{\bibfnamefont{S.}~\bibnamefont{Arcidiacono}},
  \bibinfo{author}{\bibfnamefont{J.}~\bibnamefont{Mantzaras}},
  \bibinfo{author}{\bibfnamefont{S.}~\bibnamefont{Ansumali}},
  \bibinfo{author}{\bibfnamefont{I.~V.} \bibnamefont{Karlin}},
  \bibinfo{author}{\bibfnamefont{C.}~\bibnamefont{Frouzakis}},
  \bibnamefont{and} \bibinfo{author}{\bibfnamefont{K.~B.}
  \bibnamefont{Boulouchos}}, \bibinfo{journal}{Phys. Rev. E}
  \textbf{\bibinfo{volume}{74}}, \bibinfo{pages}{056707}
  (\bibinfo{year}{2006}).

\bibitem[{\citenamefont{He et~al.}(1998)\citenamefont{He, Chen, and
  Doolen}}]{HeChenDoolen98}
\bibinfo{author}{\bibfnamefont{X.}~\bibnamefont{He}},
  \bibinfo{author}{\bibfnamefont{S.}~\bibnamefont{Chen}}, \bibnamefont{and}
  \bibinfo{author}{\bibfnamefont{G.~D.} \bibnamefont{Doolen}},
  \bibinfo{journal}{J. Comput. Phys.} \textbf{\bibinfo{volume}{146}},
  \bibinfo{pages}{282} (\bibinfo{year}{1998}).

\bibitem[{\citenamefont{Szalm\'as}(2007)}]{Szalmas07}
\bibinfo{author}{\bibfnamefont{L.}~\bibnamefont{Szalm\'as}},
  \bibinfo{journal}{Physica A}  (\bibinfo{year}{2007}).

\bibitem[{\citenamefont{Zhang et~al.}(2006)\citenamefont{Zhang, Shan, and
  Chen}}]{Zhang06}
\bibinfo{author}{\bibfnamefont{R.}~\bibnamefont{Zhang}},
  \bibinfo{author}{\bibfnamefont{X.}~\bibnamefont{Shan}}, \bibnamefont{and}
  \bibinfo{author}{\bibfnamefont{H.}~\bibnamefont{Chen}},
  \bibinfo{journal}{Phys. Rev. E} \textbf{\bibinfo{volume}{74}},
  \bibinfo{eid}{046703} (\bibinfo{year}{2006}).

\bibitem[{\citenamefont{Wilson and Kogut}(1974)}]{WilsonKogut}
\bibinfo{author}{\bibfnamefont{K.~G.} \bibnamefont{Wilson}} \bibnamefont{and}
  \bibinfo{author}{\bibfnamefont{J.~G.} \bibnamefont{Kogut}},
  \bibinfo{journal}{Phys. Rep.} \textbf{\bibinfo{volume}{12C}},
  \bibinfo{pages}{75} (\bibinfo{year}{1974}).

\bibitem[{\citenamefont{Willis}(1962)}]{Willis62}
\bibinfo{author}{\bibfnamefont{D.~R.} \bibnamefont{Willis}},
  \bibinfo{journal}{Phys. Fluids} \textbf{\bibinfo{volume}{5}},
  \bibinfo{pages}{127} (\bibinfo{year}{1962}).

\bibitem[{\citenamefont{Bird}(1994)}]{Bird}
\bibinfo{author}{\bibfnamefont{G.~A.} \bibnamefont{Bird}},
  \emph{\bibinfo{title}{Molecular {G}as {D}ynamics and the {D}irect
  {S}imulation of {G}as {F}lows}} (\bibinfo{publisher}{Clarendon Press,
  Oxford}, \bibinfo{year}{1994}).

\bibitem[{\citenamefont{Gorban and Karlin}(2005)}]{GKbook}
\bibinfo{author}{\bibfnamefont{A.~N.} \bibnamefont{Gorban}} \bibnamefont{and}
  \bibinfo{author}{\bibfnamefont{I.~V.} \bibnamefont{Karlin}},
  \emph{\bibinfo{title}{Invariant {M}anifolds for {P}hysical and {C}hemical
  {K}inetics}} (\bibinfo{publisher}{Springer, Berlin}, \bibinfo{year}{2005}).

\bibitem[{\citenamefont{Gorban and Karlin}(1996)}]{GorbanPRL96}
\bibinfo{author}{\bibfnamefont{A.~N.} \bibnamefont{Gorban}} \bibnamefont{and}
  \bibinfo{author}{\bibfnamefont{I.~V.} \bibnamefont{Karlin}},
  \bibinfo{journal}{Phys. Rev. Lett.} \textbf{\bibinfo{volume}{77}},
  \bibinfo{pages}{282} (\bibinfo{year}{1996}).

\bibitem[{\citenamefont{Karlin and Gorban}(2002)}]{Karlin02}
\bibinfo{author}{\bibfnamefont{I.~V.} \bibnamefont{Karlin}} \bibnamefont{and}
  \bibinfo{author}{\bibfnamefont{A.~N.} \bibnamefont{Gorban}},
  \bibinfo{journal}{Ann. Phys. (Leipzig)} \textbf{\bibinfo{volume}{11}},
  \bibinfo{pages}{783} (\bibinfo{year}{2002}).

\bibitem[{\citenamefont{Karlin et~al.}(1998)\citenamefont{Karlin, Gorban,
  Dukek, and Nonnenmacher}}]{Karlin98}
\bibinfo{author}{\bibfnamefont{I.~V.} \bibnamefont{Karlin}},
  \bibinfo{author}{\bibfnamefont{A.~N.} \bibnamefont{Gorban}},
  \bibinfo{author}{\bibfnamefont{G.}~\bibnamefont{Dukek}}, \bibnamefont{and}
  \bibinfo{author}{\bibfnamefont{T.~F.} \bibnamefont{Nonnenmacher}},
  \bibinfo{journal}{Phys. Rev. E} \textbf{\bibinfo{volume}{57}},
  \bibinfo{pages}{1668} (\bibinfo{year}{1998}).

\bibitem[{\citenamefont{Struchtrup and Torrilhon}(2003)}]{Struchtrup03}
\bibinfo{author}{\bibfnamefont{H.}~\bibnamefont{Struchtrup}} \bibnamefont{and}
  \bibinfo{author}{\bibfnamefont{M.}~\bibnamefont{Torrilhon}},
  \bibinfo{journal}{Phys. Fluids} \textbf{\bibinfo{volume}{15}},
  \bibinfo{pages}{2668} (\bibinfo{year}{2003}).

\bibitem[{\citenamefont{Chen et~al.}(2003)\citenamefont{Chen, Kandasamy,
  Orszag, Shock, Succi, and Yakhot}}]{SCI}
\bibinfo{author}{\bibfnamefont{H.}~\bibnamefont{Chen}},
  \bibinfo{author}{\bibfnamefont{S.}~\bibnamefont{Kandasamy}},
  \bibinfo{author}{\bibfnamefont{S.}~\bibnamefont{Orszag}},
  \bibinfo{author}{\bibfnamefont{R.}~\bibnamefont{Shock}},
  \bibinfo{author}{\bibfnamefont{S.}~\bibnamefont{Succi}}, \bibnamefont{and}
  \bibinfo{author}{\bibfnamefont{V.}~\bibnamefont{Yakhot}},
  \bibinfo{journal}{Science} \textbf{\bibinfo{volume}{301}},
  \bibinfo{pages}{633} (\bibinfo{year}{2003}).

\bibitem[{\citenamefont{Ansumali et~al.}(2004)\citenamefont{Ansumali, Karlin,
  and Succi}}]{AK8}
\bibinfo{author}{\bibfnamefont{S.}~\bibnamefont{Ansumali}},
  \bibinfo{author}{\bibfnamefont{I.~V.} \bibnamefont{Karlin}},
  \bibnamefont{and} \bibinfo{author}{\bibfnamefont{S.}~\bibnamefont{Succi}},
  \bibinfo{journal}{Physica A} \textbf{\bibinfo{volume}{338}},
  \bibinfo{pages}{379} (\bibinfo{year}{2004}).

\end{thebibliography}
\end{document}